\documentclass[aps,prb,twocolumn,superscriptaddress,showpacs]{revtex4}
\usepackage{graphics}
\usepackage{graphicx}
\usepackage[english]{babel}
\usepackage{color}

\begin{document}

\title{Size-induced depression of first-order
transition lines and entropy-jump in extremely-layered
nanocrystalline vortex matter}

\author{M.I. Dolz}
\affiliation{Departamento de F\'{i}sica, Universidad Nacional de
San Luis and CONICET,  San Luis, Argentina}

\author{Y. Fasano}
\affiliation{Low Temperatures Division, Centro At\'{o}mico
Bariloche, CNEA, Bariloche, Argentina}

\author{N. R. Cejas Bolecek}
\affiliation{Low Temperatures Division, Centro At\'{o}mico
Bariloche, CNEA,  Bariloche, Argentina}

\author{H. Pastoriza}
\affiliation{Low Temperatures Division, Centro At\'{o}mico
Bariloche, CNEA,  Bariloche, Argentina}

\author{V. Mosser}
\affiliation{Itron SAS, F-92448 Issy-les-Moulineaux, France}

\author{M. Li}
\affiliation{Kamerlingh Onnes Laboratorium, Rijksuniversiteit
Leiden, 2300 RA Leiden, The Netherlands}

\author{M. Konczykowski}
\affiliation{Laboratoire des Solides Irradi\'{e}s, Ecole
Polytechnique, CNRS URA-1380, 91128 Palaiseau, France}

\date{\today}

\begin{abstract}
We detect the persistence of the  solidification and
order-disorder first-order transition lines in the phase diagram
of nanocrystalline Bi$_{2}$Sr$_{2}$CaCu$_{2}$O$_{8}$ vortex matter
down to a system size of less than hundred vortices.  The
temperature-location of the vortex solidification transition line
is not altered by decreasing the sample size although there is a
depletion of the entropy-jump at the transition with respect to
macroscopic vortex matter. The solid order-disorder phase
transition field  moves upward on decreasing the system size due
to the increase of the surface-to-volume ratio of vortices
entailing a decrease on the average vortex binding energy.
\end{abstract}

 \keywords{}

\maketitle

Understanding size-induced structural phase transitions and
size-depression of transition temperatures is  crucial for the
development of devices made of nanoscale building-blocks of
functional materials. These blocks can be as small as nanocrystals
with hundred particles or less. In the case of hard condensed
matter, the physical properties and stable phases of nanocrystals
strongly depend on its size and shape. For instance, the
size-induced depression of characteristic temperatures and
thermodynamic properties in
melting,~\cite{Coombes1972,Goldstein1992}, solid-solid
first-order,~\cite{Tolbert1994} ferromagnetic and
superconducting~\cite{Guisbiers2009} transitions has been the
subject of  experimental and theoretical studies in a variety of
systems such as metallic nanoparticles and films, and
semiconductor nanocrystals. Typically, the thermodynamic
properties at  phase transitions, such as entropy, enthalpy, and
transition temperature, decrease when reducing the nanocrystal
size.~\cite{Coombes1972,Goldstein1992,Tolbert1994,Guisbiers2009}
This trend is interpreted as a result of the high proportion of
particles located at the surface of the nanocrystal that have a
lesser binding energy than those of the volume.

Vortex matter in high-temperature superconductors is a rich
soft-condensed matter playground where relevant questions on this
issue can be easily answered.  The number of particles (vortices)
can be tuned by varying the applied fields since the inter-vortex
spacing $a \propto 1/B$.~\cite{Blatter1994} Therefore
nanocrystalline vortex matter can be nucleated by conveniently
reducing the applied field and sample size down to
 tens of microns by means of lithographic engineering techniques. The binding energy of vortex matter
can be selected by changing $H$ and superconducting material since
it is proportional to the inter-vortex interaction energy that
depends on the $a/\lambda$ ratio, with $\lambda$ the material
penetration depth. In addition, the location  of  thermodynamic
phase transitions in the density-temperature vortex phase diagram
depends on the balance between three energy scales: thermal,
inter-vortex and vortex-disorder interactions.~\cite{Blatter1994}
This balance can be tuned by temperature and choosing
superconducting materials with different disorder (pinning) and
vortex elastic properties. In extremely-anisotropic high-$T_{\rm
c}$'s this energy-competition is fostered by the softening of the
vortex lattice induced by the material
layerness.~\cite{Blatter1994} The phase diagram of macroscopic
samples is dominated by a liquid-solid first-order transition
(FOT) at which the thermal energy overwhelms the binding energy.
The ordered solid vortex phase presents a structural FOT  to a
disordered vortex glass on increasing $H$.~\cite{Blatter1994}
Studying the persistence or disappearance of this melting and
solid-solid transitions on decreasing the number of vortices allow
us to test the validity of the thermodynamic limit in FOTs.

A model high-$T_{\rm c}$ material to study these issues is the
layered Bi$_{2}$Sr$_{2}$CaCu$_{2}$O$_{8}$ compound that in its
pristine form is a rather clean system. The FOT line $H_{\rm FOT}$
in this material~\cite{Pastoriza94a,Zeldov95a} separates a solid
vortex phase at low fields with a liquid~\cite{Nelson1988} or
decoupled gas~\cite{Glazman1991} of pancake vortices with reduced
shear viscosity~\cite{Pastoriza1995} in the high-temperature
range. At low temperatures, the so-called second-peak, $H_{\rm
SP}$, order-disorder FOT separates the vortex solid from a glassy
vortex phase at high $H$.~\cite{Avraham2001} Direct imaging of
vortices in pristine Bi$_{2}$Sr$_{2}$CaCu$_{2}$O$_{8}$ reveals the
low-field vortex solid has quasi long-range positional
order.~\cite{Fasano1999} This phase is magnetically irreversible
due to the effect of bulk pinning and surface barriers that
dominate at different temperature and measuring-time
ranges.~\cite{Chikumoto1992}

The persistence of these transition lines and the evolution of
their thermodynamic properties in nanocrystalline vortex matter is
still open to discussion. In the case of
Bi$_{2}$Sr$_{2}$CaCu$_{2}$O$_{8}$, Josephson-plasma-resonance data
suggest that the $H_{\rm FOT}$ transition entails a single-vortex
decoupling process between stacks of pancake vortices in adjacent
CuO planes.~\cite{Colson2003}  This can be further tested by
decreasing the system size to less than hundred vortices. This
issue has not been studied probably since achieving the required
experimental $B$-resolution is not straightforward. Regarding the
$H_{\rm SP}$ FOT, there is a report on its disappearance for a
system with hundreds vortices.~\cite{Wang2000} This was explained
invoking bulk currents arguments in spite of their two-quadrant dc
magnetization loops being dominated by surface-barrier
physics.~\cite{Wang2000}

\begin{figure}[ttt]
\includegraphics[width=\columnwidth,angle=0]{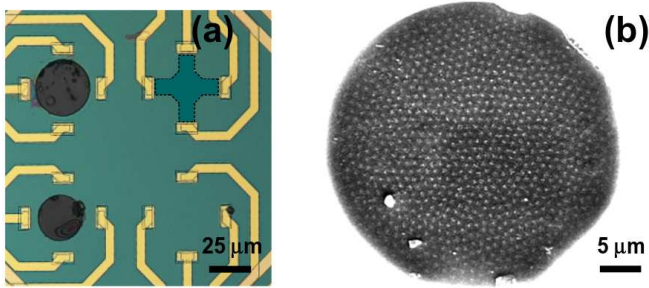}
\caption{(a) Micron-sized Bi$_{2}$Sr$_{2}$CaCu$_{2}$O$_{8}$ disks
 mounted on the  Hall sensors with $16 \times
16$\,$\mu$m$^{2}$ active surfaces (indicated with dotted lines).
(b) Nanocrystalline vortex solid imaged by magnetic decoration at
20\,Oe in a 40\,$\mu$m diameter disk. \label{figure1}}
\end{figure}

We studied the effect of confinement on the location of $H_{\rm
FOT}$ and $H_{\rm SP}$ and on their associated thermodynamic
properties by nucleating nanocrystalline vortex matter in
micron-sized disks of Bi$_{2}$Sr$_{2}$CaCu$_{2}$O$_{8}$
single-crystals ($T_{\rm c}=90$\,K). The disks, engineered by
optical lithography and physical ion-milling,~\cite{Dolz14a} have
thicknesses of $\sim 1$\,$\mu$m and diameters between 20 to
50\,$\mu$m. Magnetic decoration~\cite{Fasano1999} experiments show
the penetration of regular vortex structures as in
Refs.\,\onlinecite{Dolz14b,Cejas2014}, see Fig.\,\ref{figure1}
(b). Vortices are observed at distances smaller than $1\,\mu$m
from the edge indicating no degradation of the sample edge as a
result of the preparation method.

 The local magnetization  of the disks, $H_{\rm s}$, was measured using 2D electron gas Hall sensors with $16 \times
16$\,$\mu$m$^{2}$ active surfaces.  Figure \ref{figure1} (a) shows
two of the studied disks glued onto the sensors with Apiezon N
grease  to improve thermal contact. A  local coil generates an ac
field $H_{\rm ac}$ parallel to the dc one, with amplitudes ranging
$1-3$\,Oe rms and frequencies between 1 to 200\,Hz. dc and ac
local magnetic measurements were performed using the same set-up
as a function of temperature, magnetic field, and frequency. In
the ac measurements we simultaneously acquire the in- and
out-of-phase components of the fundamental and the third-harmonics
of the Hall voltage using a digital-signal-processing lock-in
(EG\&G 7265). The in-phase component of the first-harmonic signal,
$B'$, is used to calculate the transmittivity of the vortex system
$T'= [B'(T) - B'(T \ll T_{\rm c})]/[B'(T>T_{\rm c}) - B'(T \ll
T_{\rm c})]$.~\cite{Gilchrist1993} This magnitude is more
sensitive to discontinuities in the local induction than direct
measurements of the static  $B$. The modulus of the normalized
third harmonic signal $\mid T_{\rm h3} \mid$~\cite{Gilchrist1993}
is non-negligible when the magnetic response becomes non-linear.
Our ac  data have a resolution of 5\,mG, one order of magnitude
better than dc measurements.

\begin{figure}[ttt]
\includegraphics[width=0.92\columnwidth,angle=0]{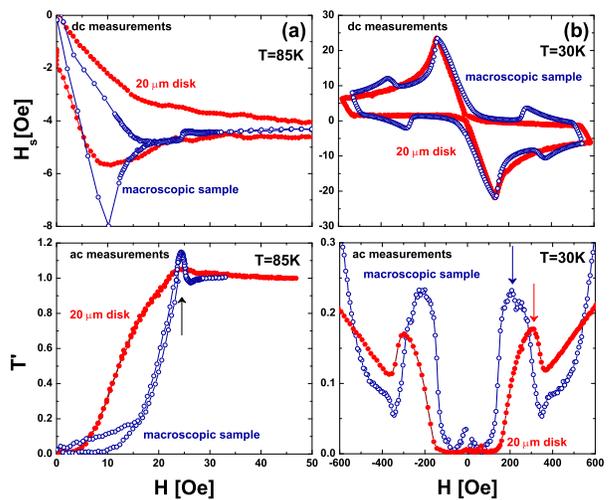}
\caption{Magnetic response of nanocrystalline and macroscopic
Bi$_{2}$Sr$_{2}$CaCu$_{2}$O$_{8}$ vortex matter. (a) The  $H_{\rm
FOT}$ transition  is detected as a jump in the dc sample field
(top panel) and as a peak in the ac transmittivity (bottom panel).
(b) The order-disorder $H_{\rm SP}$ field is taken at the onset of
the increase in critical current, see arrows. \label{figure2}}
\end{figure}

Figure \,\ref{figure2} contrasts the magnetic response of the
smallest measured 20\,$\mu$m disk with that of the macroscopic
sample from which the disks were engineered. For the disks the
detection of the $B$-jump at $H_{\rm FOT}$ is elusive in dc loops,
see top panel of Fig.\,\ref{figure2}\,(a). Performing ac $T'(H)$
measurements allows determining $H_{\rm FOT}$ in the disks from
the field-location of the paramagnetic peak entailed at this FOT,
see arrow in Fig.\,\ref{figure2}\,(a). The paramagnetic peaks are
located at the same field for the macroscopic sample and the
disks.  The nearly-vanishing remanent magnetization and the
two-quadrant locus of the dc loops of the disks indicate a larger
influence of surface barriers for vortex flux entry/exit.
Performing ac magnetometry is then imperative to have access to
the faster decaying bulk currents emerging from the
surface-barrier background \cite{Chikumoto1992} and therefore to
properly track FOT lines in nanocrystalline vortex matter.

\begin{figure}[ttt]
\includegraphics[width=\columnwidth,angle=0]{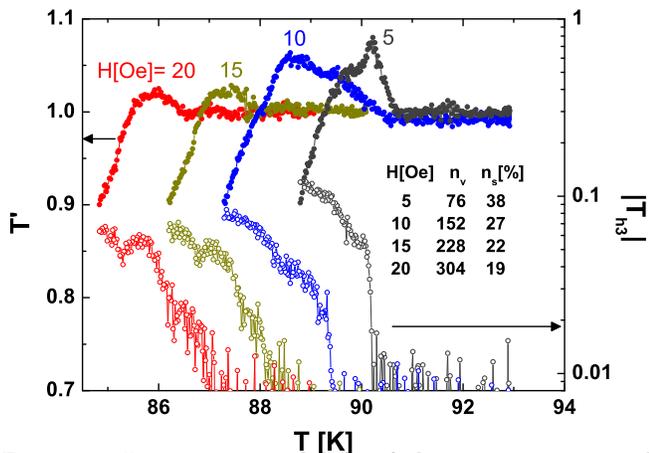}
\vspace{-1 cm}
 \caption{Temperature-evolution of the
transmittivity and third-harmonic for the 20\,$\mu$m diameter
disk. For each measurement $H$ we show the number of vortices
nucleated in the sample, $n_{\rm v}$, and the surface-to-volume
vortex ratio, $n_{\rm s}$. \label{figure3}}
\end{figure}

\begin{figure}[ttt]
\includegraphics[width=\columnwidth,angle=0]{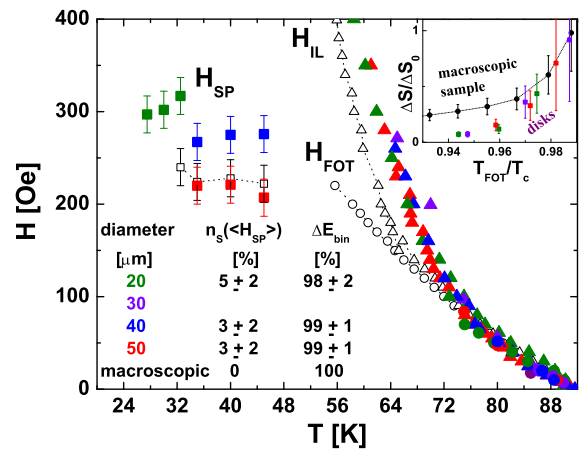}
\vspace{-0.7 cm}
 \caption{Phase diagram of  nanocrystalline and
macroscopic~\cite{Dolz14a} Bi$_{2}$Sr$_{2}$CaCu$_{2}$O$_{8}$
vortex matter indicating the surface-to-volume vortex ratio for
the average $H_{\rm SP}$, $n_{\rm s}(\,<\,H_{\rm SP}\,>\,)$, and
the nanocrystal-to-macroscopic bindig energy ratio, $\Delta E_{\rm
bin}$. Insert: Normalized entropy-jump at $H_{\rm FOT}$.
\label{figure4}}
\end{figure}

Figure\,\ref{figure3} shows $T'(T)$ and $\mid T_{\rm h3}(T) \mid$
data for the 20\,$\mu$m diameter disk for low number of vortices,
$76<n_{\rm v}<304$. The paramagnetic peak fingerprinting the
$H_{\rm FOT}$ is clearly visible for a nanocrystal of only 76
vortices (5\,Oe). This peak shifts to lower temperatures on
increasing $H$ and echoes in the $\mid T_{\rm h3}(T) \mid$ signal,
indicating the high-resolution of our ac measurements. Similar
behavior in $T'(T)$ and $\mid T_{\rm h3}(T) \mid$ are also
observed  for the 30, 40, and 50\,$\mu$m-diameter disks. These
results are summarized in the vortex phase diagram of
Fig.\,\ref{figure4}. The $H_{\rm FOT}$ transition line  is
obtained from tracking the paramagnetic peak in the
temperature-field plane. For $T < 75\,$K no paramagnetic peak is
detected, presumably due to a masking effect produced by enhanced
surface barriers in small samples. In the case of the disks, the
$H_{\rm FOT}$ line merges that of the macroscopic reference
sample.

Therefore, the $H_{\rm FOT}$
 transition persists and no melting-point depression is
observed even for a nanocrystal with less than hundred vortices.
Ordinary melting is the result of thermal breaking of the
inter-particle bonds. In nanocrystals the average binding energy,
$<E_{\rm bin}>$, is smaller  since the considerable fraction of
particles located at the surface, $n_{\rm s}$, has a lesser
binding energy. The ratio between the
nanocrystalline-to-macroscopic $<E_{\rm bin}>$ can be obtained in
a rough estimation as $\Delta E_{\rm bin}= (0.5 n_{\rm s}) + (1 -
n_{\rm s})$, the first term coming from considering that the
binding energy of vortices of the outer shell  is reduced to half
the value for vortices at the interior of the sample, considered
in the second term.  For instance, for the smallest measured
nanocrystal $n_{\rm s}= 38$\,\% and $\Delta E_{\rm bin}\sim
80\,\%$, a decrease of $<E_{\rm bin}>$ that in hard condensed
matter should be enough in order to produce a noticeable
melting-point depression.~\cite{Goldstein1992} \textit{Since for
nanocrystalline Bi$_{2}$Sr$_{2}$CaCu$_{2}$O$_{8}$ vortex matter
this is not observed, the symmetry broken at $H_{\rm FOT}$ should
be other than the in-plane inter-vortex bonds}. There is evidence
that in Bi$_{2}$Sr$_{2}$CaCu$_{2}$O$_{8}$ the $H_{\rm FOT}$
transition occurs concomitantly with a c-axis decoupling between
pancake vortices of adjacent CuO planes.~\cite{Colson2003} For the
smallest system we studied the number of pancake vortices is
larger than 50.000. Considering this, our results suggest that the
broken symmetry on cooling at $H_{\rm FOT}$ is the establishment
of the c-axis superconducting phase coherence produced by the
coupling of pancake vortices in the whole sample thickness, rather
than an improvement of the in-plane structural order of vortex
matter exclusively. Indeed, in-plane disordered vortex structures
are nucleated in irradiated samples that still present a $H_{\rm
FOT}$ transition.~\cite{Menghini2003}

Other thermodynamic properties rather than $T_{\rm FOT}$ are
however affected by decreasing the sample size. The insert to
Fig.\,\ref{figure4} shows the evolution of the normalized
entropy-jump, $\Delta S / \Delta S_{\rm 0}$, at the liquid-solid
transition  for nanocrystalline and macroscopic
Bi$_{2}$Sr$_{2}$CaCu$_{2}$O$_{8}$ vortex matter. The entropy-jump
at the transition, $\Delta S$, is shown normalized by the $\Delta
S_{\rm 0}$ value extrapolated at zero field. The $\Delta S$ versus
reduced temperature $T_{FOT}/T_{\rm c}$ data were obtained from
applying the Clausius-Clapeyron relation, $\Delta S \propto
(\Delta B/B_{\rm FOT})dH_{\rm FOT}/dT$,~\cite{Dolz14a} with
$\Delta B$ the enthalpy-jump. In the case of the disks the
enthalpy was estimated from $T'(T)$ data considering that at the
transition $T'= 1+ (2 \Delta B)/(\pi H_{\rm ac})$ \cite{Dolz14a}.
The data for the disks with 50 to 20\,$\mu$m diameter are packed
in a single trend of $\Delta S / \Delta S_{\rm 0}$ lying below the
 macroscopic sample data.

Since the $T_{\rm FOT}$ does not vary with the system size in the
range studied, this decrease in $\Delta S$ comes from a decrease
on the enthalpy entailed in the transition for nanocrystalline
vortex systems. Therefore, irrespective of the system size, vortex
matter undergoes a solidification transition at the same $T_{\rm
FOT}$ entailing the same type of symmetry breaking. However,
assuming that the entropy of the liquid is independent of the
system size, the solid nanocrystalline vortex matter has a larger
entropy than macroscopic crystals. In previous works we reported
that there is an increase of the density of topological defects in
nanocrystalline vortex matter with respect to the macroscopic
case.~\cite{Cejas2014} Therefore the sample-size-induced
enthalpy-jump decrease in the $H_{\rm FOT}$ transition is in
agreement with a worsening of the positional in-plane structural
order induced by confinement effects.

In order to better understand these findings, theoretical work on
the sample-size dependence of the entropy-jump at the FOT and the
electromagnetic-to-Josephson interactions ratio is required. Any
attempt of quantitative explanation of this evolution should
include a realistic model on the evolution of the  binding energy
on decreasing the system size for  extremely-layered vortex
matter.

We also study the effect of decreasing the vortex number on the
solid order-disorder $H_{\rm SP}$ transition.
Figure\,\ref{figure2} (b) compares dc and ac results: While the dc
loops do not show evidence of $H_{\rm SP}$ for the 20\,$\mu$m
disk, local maximum and minimum in the ac data are evident. These
features come from the increase in critical current entailed in
the $H_{\rm SP}$ transition, detected by applying the ac technique
more sensitive to bulk current contributions. Therefore the
previously reported absence of the $H_{\rm SP}$ transition for
micron-sized Bi$_{2}$Sr$_{2}$CaCu$_{2}$O$_{8}$
samples~\cite{Wang2000} comes from technical limitations of dc
magnetic techniques for which surface currents play a determining
role in screening.~\cite{Chikumoto1992}

In order to quantify the effect of decreasing the number of
vortices on the solid order-disorder transition we tracked this
line considering the onset of the effect associating $H_{\rm SP}$
to the local maximum in $T'(H)$. Figure\,\ref{figure4} shows a
remarkable result: On decreasing the sample size   $H_{\rm SP}$
monotonically moves up in $H$ increasing the field-stability  of
the ordered low-field vortex-solid phase.  The $H_{\rm SP}$ line,
as any structural solid-to-solid FOT, entails a small
discontinuity in the binding energy that is smaller for the
disordered phase, the high-field vortex-glass. At a given field,
when decreasing the sample size $n_{\rm s}$ is enhanced and then
the $<E_{\rm bin}>$, and accordingly $\Delta E_{\rm bin}$, are
smaller than for a larger nanocrystal. The fraction of
surface-to-volume vortices at the average value of the almost
temperature-independent $H_{\rm SP}$, $n_{\rm s}(<H_{\rm SP}>)$,
is of the order of $4 \pm 2\,\%$ for all the studied micron-sized
disks (see Fig.\,\ref{figure4}). Considering this information, and
our rough estimation for the decrease in the $<E_{\rm bin}>$ of
vortex nanocrystals,  $\Delta E_{\rm bin} \sim (98 \pm 2)\,\%$ at
the average $H_{\rm SP}$, independently of the sample size. The
order-disorder transition condition is that at $H_{\rm SP}$ the
elastic energy of the vortex system, proportional to its binding
energy, is balanced by the interaction energy of vortices with the
disorder introduced by pinning.~\cite{Blatter1994} Since no
dramatic changes in the pinning energy are introduced by our
method of decreasing the sample size as observed in the dc loops
of Fig.\,\ref{figure2},~\cite{MO} in order to reach this
energy-balance condition when decreasing the sample size, $H$ has
to be increased such to attain the same interaction energy
$\propto \Delta E_{\rm bin}$.

The non-thermodynamic line at which magnetic response becomes
irreversible, $H_{\rm IL}$, is also affected by confinement. For
nanocrystalline vortex matter this line is located at higher
fields than for the macroscopic case, see Fig.\,\ref{figure4}.
This  might have origin in the different aspect-ratio of the
 samples, and in their probably dissimilar surface roughness.  In the high-temperature range the $H_{\rm IL}$ lines
for nanocrystalline and macroscopic vortex matter merge into a
single bunch of data with the $H_{\rm FOT}$ line. At high fields,
the onset of $\mid T_{\rm h3} \mid$ develops before the
paramagnetic peak on cooling, see Fig.\,\ref{figure3}, indicating
the existence of a narrow non-linear vortex region spanning at
$H>H_{\rm FOT}$.~\cite{Indenbom1996} This phase-region might have
origin on a residual effect of pinning~\cite{Vinokur1991}, or
Bean-Livingston barriers~\cite{Fuchs1998} on the liquid phase.
This phenomenology was already detected  in macroscopic samples,
although spanning a larger phase region than in vortex
nanocrystals.~\cite{Dolz14a}

In conclusion, the liquid-to-solid FOT remains robust and persists
at the same location even when decreasing the system size  to less
than hundred vortices. This supports the scenario of the FOT line
being a single-vortex decoupling transition~\cite{Colson2003} that
depends, at best, on the density of the surrounding vortex matter.
The entropy-jump entailed in the transition decreases for
nanocrystalline vortex matter in agreement with direct-imaging
experiments revealing a worsening of the positional order with
respect to the macroscopic case.~\cite{Cejas2014} The solid
order-disorder FOT at $H_{\rm SP}$ also persists on decreasing the
system size down to 20\,$\mu$m contrary to what has been
reported.~\cite{Wang2000} The $H_{\rm SP}$ moves upward on
reducing the sample size such as the average binding energy of the
nanocrystal, and thus its interaction energy, remains roughly the
same. The identification of these FOTs increases in difficulty
when decreasing the sample size due to the predominance of
surface-barrier-related currents and therefore applying ac
magnetometry is mandatory in order to improve the sensitivity to
bulk currents.

\end{document}